\documentclass[12pt]{article}
\usepackage{fixltx2e}
\usepackage{textcomp}
\usepackage{fullpage}
\usepackage{graphicx}
\usepackage[english]{babel}
\usepackage{pifont}
\usepackage{color}
\usepackage[usenames,dvipsnames]{xcolor}
\usepackage{amsfonts}
\usepackage{longtable}
\usepackage{indentfirst}
\usepackage{setspace}
\usepackage{verbatim}
\usepackage{upgreek}
\usepackage{mathrsfs}
\usepackage{amsmath}
\usepackage{amsthm}
\usepackage{amssymb}
\usepackage[normalem]{ulem}
\usepackage{booktabs}
\usepackage{natbib}
\usepackage{float}
\usepackage{latexsym}
\raggedright
\definecolor{citescol}{RGB}{73,0,165}
\definecolor{urlscol}{RGB}{0,107,124}
\definecolor{linkscol}{RGB}{187,24,0}

\definecolor{tahcommentcol}{RGB}{185,88,187}
\definecolor{frcommentcol}{RGB}{0,255,0}
\definecolor{shcommentcol}{RGB}{0,0,255}

\DeclareMathAlphabet{\msfsl}{T1}{cmr}{m}{it}
\DeclareMathAlphabet{\msyf}{OMX}{pcr}{m}{it}

\begin{document}
%%%%%%%%%%%%%%%%%%%%%%%%%%%%%%%%%%%%
%  TITLE PAGE  
%%%%%%%%%%%%%%%%%%%%%%%%%%%%%%%%%%%%
\begin{flushright}
Version dated: \today
\end{flushright}
\bigskip

\begin{center}

\title{\large The inference of gene trees with species trees}

%\maketitle

\begin{center}

\noindent{\Large \bf The inference of gene trees with species trees}
\medskip
\end{center}

\noindent {\normalsize \sc Gergely J. Sz\"oll\H{o}si$^1$, Eric Tannier$^{2,3,4}$, Vincent Daubin$^{2,3}$, Bastien Boussau$^{2,3}$}\\
\medskip

\noindent {\small \it 
$^1$ELTE-MTA ``Lend\"ulet'' Biophysics Research Group, P\'azm\'any
P. stny. 1A., 1117 Budapest, Hungary;\\
$^2$Laboratoire de Biom\'etrie et Biologie Evolutive, Centre National de la Recherche Scientifique, Unit\'e Mixte de Recherche 5558, Universit\'e Lyon 1, F-69622 Villeurbanne, France;\\
$^3$Universit\'e de Lyon, F-69000 Lyon, France;\\
$^4$Institut National de Recherche en Informatique et en Automatique Rh\^one-Alpes, F-38334 Montbonnot, France;}\\
\end{center}

\medskip
\noindent{\bf Corresponding author:} Bastien Boussau, Laboratoire de Biom\'etrie et Biologie Evolutive, Centre National de la Recherche Scientifique, Unit\'e Mixte de Recherche 5558, Universit\'e Lyon 1, F-69622 Villeurbanne, France; Universit\'e de Lyon, F-69000 Lyon, France;  E-mail: bastien.boussau@univ-lyon1.fr\\

\newpage

%\email[]{bastien.boussau@univ-lyon1.fr}

%\author{Gergely J. Sz\"oll\H{o}si}
%%\email[]{ssolo@elte.hu}
%\affiliation{ELTE-MTA ``Lend\"ulet'' Biophysics Research Group
%1117 Bp., P\'azm\'any P. stny. 1A., Budapest, Hungary;}
%\author{Eric Tannier}
%\affiliation{Laboratoire de Biom\'etrie et Biologie Evolutive, Centre National de la Recherche Scientifique, Unit\'e Mixte de Recherche 5558, Universit\'e Lyon 1, F-69622 Villeurbanne, France;}
%\affiliation{Universit\'e de Lyon, F-69000 Lyon, France;}
%\affiliation{Institut National de Recherche en Informatique et en Automatique Rh\^one-Alpes, F-38334 Montbonnot, France}
%\author{Vincent Daubin}
%\affiliation{Laboratoire de Biom\'etrie et Biologie Evolutive, Centre National de la Recherche Scientifique, Unit\'e Mixte de Recherche 5558, Universit\'e Lyon 1, F-69622 Villeurbanne, France;}
%\affiliation{Universit\'e de Lyon, F-69000 Lyon, France;}
%\author{Bastien Boussau}
%\affiliation{Laboratoire de Biom\'etrie et Biologie Evolutive, Centre National de la Recherche Scientifique, Unit\'e Mixte de Recherche 5558, Universit\'e Lyon 1, F-69622 Villeurbanne, France;}
%\affiliation{Universit\'e Claude Bernard Lyon 1 UMR CNRS 5558 - LBBE, Lyon, France.}
%\affiliation{Universit\'e de Lyon, F-69000 Lyon, France;}

\begin{abstract}
Molecular phylogeny has focused mainly on improving models for the reconstruction of gene trees based on sequence alignments. Yet, most phylogeneticists seek to reveal the history of species. Although the histories of genes and species are tightly linked, they are seldom identical, because genes duplicate, are lost or horizontally transferred, and because alleles can co-exist in populations for periods that may span several speciation events. Building models describing the relationship between gene and species trees can thus improve the reconstruction of gene trees when a species tree is known, and vice-versa. Several approaches have been proposed to solve the problem in one direction or the other, but in general neither gene trees nor species trees are known. Only a few studies have attempted to jointly infer gene trees and species trees. In this article we review the various models that have been used to describe the relationship between gene trees and species trees. These models account for gene duplication and loss, transfer or incomplete lineage sorting. Some of them consider several types of events together, but none exists currently that considers the full repertoire of processes that generate gene trees along the species tree. Simulations as well as empirical studies on genomic data show that combining gene tree-species tree models with models of sequence evolution improves gene tree reconstruction. In turn, these better gene trees provide a better basis for studying genome evolution or reconstructing ancestral chromosomes and ancestral gene sequences. We predict that gene tree-species tree methods that can deal with genomic data sets will be instrumental to advancing our understanding of genomic evolution.

\end{abstract}

%\keywords{gene tree reconstruction, gene tree reconciliation, incomplete lineage sorting, lateral gene transfer, phylogeny}

\section*{Introduction}

During the last fifty years, phylogeny has become more and more based on molecular data, increasingly favoring homologous sequences over morphological characters. This approach has been extremely fruitful, producing constant improvement in the accuracy and resolution of phylogenetic reconstruction together with our understanding of evolutionary processes at the molecular level. However, we have known all along that we are barking up the wrong trees: with increasing sophistication in the models of sequence evolution, we have been reconstructing trees describing the history of fragments of genomic sequence, which we will liberally call "gene" in this review, but never the history of species. Gene trees are not species trees \citep{Maddison1997}. 

Each gene tree reflects a unique story, which is linked to species history, but often significantly differs from it \citep{Szollosi:2012uq}. Gene trees reflect the process of replication at a local level: a copy of a gene at a locus in the genome, \textit{e.g.} a protein coding gene, replicates and its copies are passed on from parent to offspring, generating branching points in the gene tree. Because each gene copy has a single ancestral copy, barring recombination, all gene trees would be identical. Recombination, however, breaks up the genomic history into a series of partially independent stories, \textit{i.e.} into gene trees along the genomes of species. 

Starting from an individual site in a genome up to the species level, a hierarchy of evolutionary processes generate genomic sequences. Individual sites evolve as a result of point mutations. The fate of individuals carrying each mutation is played out at the population level, and determines whether a mutation is fixed in the population as a substitution, or is ultimately lost. The birth and death of larger stretches of sequence, \textit{e.g.} of entire genes, occurs as a result of insertions and deletions in individual genomes, the fate of which, similar to point mutations, is played out at the population level. The source of the inserted sequence differentiates between duplication events, wherein a sequence from the same genome is inserted, from lateral transfer events, wherein a sequence from an external source is inserted. Finally, species, \textit{i.e.}, populations of genomes, evolve through speciation and extinction events.

\begin{center}
	\begin{figure}[P]
	\centering
		\includegraphics[width=1.\columnwidth]{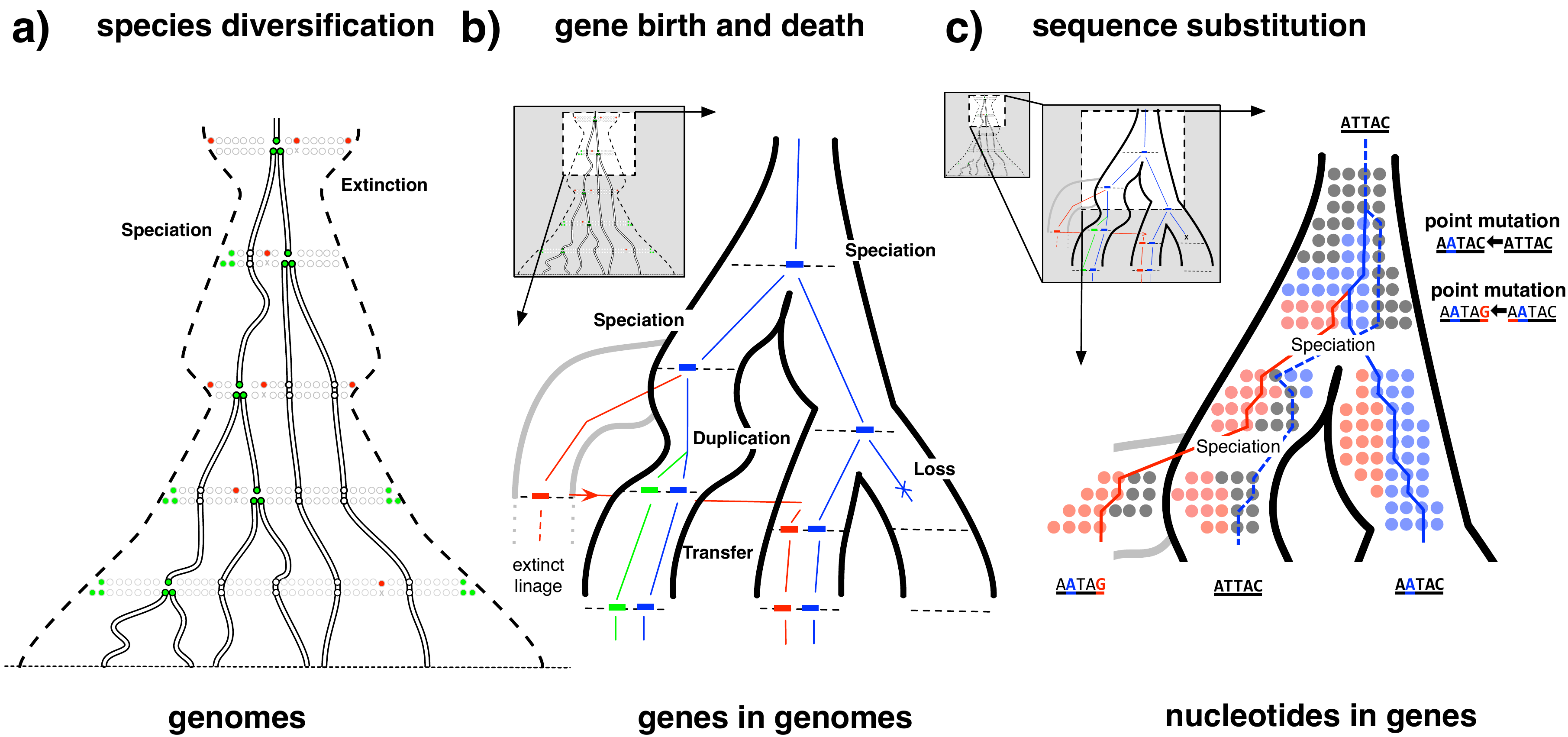}
		\caption{A hierarchy of evolutionary processes contribute to sequence evolution. a) Individual species and their genomes evolve among a population of species, according to a diversification process governing speciation and extinction events. b) Inside each genome, each gene evolves according to gene duplication, loss and transfer events. c) Individual sites evolve through point mutations. Processes at the gene and site level are played out at the population level, where changes fix or are lost. }
		\label{fig:figure1}
	\end{figure}
\end{center}

As illustrated in Fig. \ref{fig:figure1} each level of the hierarchy contributes to generating differences between gene trees. Segregating mutations that cross speciation events (a process called incomplete lineage sorting) leave topological signatures in gene trees (see Fig. \ref{fig:figure1}.c.). Current estimates indicate that up to $30\%$ of the sequence of the human genome is more closely related to Gorilla than to Chimpanzee due to this process \citep{Scally2012}. Duplication, transfer and loss events (Fig. \ref{fig:figure1}.b.) lead to large differences in both the size and phylogenetic distribution of families of homologous genes, and at the same time produce patent phylogenetic discord \citep{Szollosi:2012uq}. Finally, the species diversification processes influence lateral transfer, as most transfer events come from donor species that have gone extinct or have not been sampled \citep{Szollosi2013}.       

When reconstructing a gene tree it is desirable to integrate these different types of information in order to maximize the amount of information used and to insure the consistency of our prediction. In a pioneering attempt to do so, \cite{Goodman1979} proposed to reconstruct the history of a gene by searching for the tree which minimizes the sum of the number of nucleotide substitutions, duplications and losses. This parsimonious approach is clear and conceptually straight-forward, but it raises the difficulty of determining the relative weights of events that are very different in nature. However, if we can construct a coherent and principled approach to combine these events, it becomes possible to envisage the reconstruction of a gene tree given a sequence alignment and a known species tree. Moreover, given a set of gene alignments, it would allow us to reconstruct the species history that has most likely generated them. In a probabilistic framework, the combination of a model of sequence evolution and a model of gene family evolution accounting for duplication, loss, lateral transfer and/or incomplete lineage sorting can be based on a quite natural hypothesis of conditional independence: if we assume that the species tree is independent from the sequence conditional on the gene tree, then the probabilities of the models can be multiplied.
In addition, parsimony approaches often generate a large number of equivalent solutions and don't allow an efficient exploration or integration over solution spaces. Probabilistic models allow integrating or sampling over the very large numbes of possible scenarios. For example it is possible to estimate the probability of a particular gene tree, which is not only the probability of the most likely scenario of sequence evolution and DTL events, but the sum of the probabilities of all possible scenarios that could have generated this tree. 

In the past 15 years, several such methods, which model consistently the dependence between gene trees and the species tree, have been developed and have shown improved accuracy for inferring both gene trees and the species tree. In this review we present these methods, explain the assumptions they make, introduce how they work, and highlight some of the results obtained with them. We focus on probabilistic models, but discuss parsimony-based approaches in situations where probabilistic models have not been developed yet.

\section*{Modeling the dependence between gene tree and species tree}

%	The dependence between gene trees and the species tree needs to be put into a probabilistic model before statistical inference of a gene tree, of the species tree, or of both is performed. 

The processes known to contribute to gene family evolution include incomplete lineage sorting (ILS),  gene duplication and loss (DL), and gene transfer (T). Hybridization can be seen as a special type of transfer, affecting a large portion of the genomes, and resulting in a gene replacement in the receiving species. Allopolyploidization is a particular type of hybridization, in which the two genomes keep cohabiting in subsequent generations.
For each single process, there are published models accounting for its effect, and recently some tend to integrate several of them. So far, no model has been published that deals with all processes together in a coherent statistical framework.
	
\subsection*{Gene birth-death generates gene trees along the species tree}
\begin{center}
	\begin{figure}[P]
	\centering
		\includegraphics[width=1.\columnwidth]{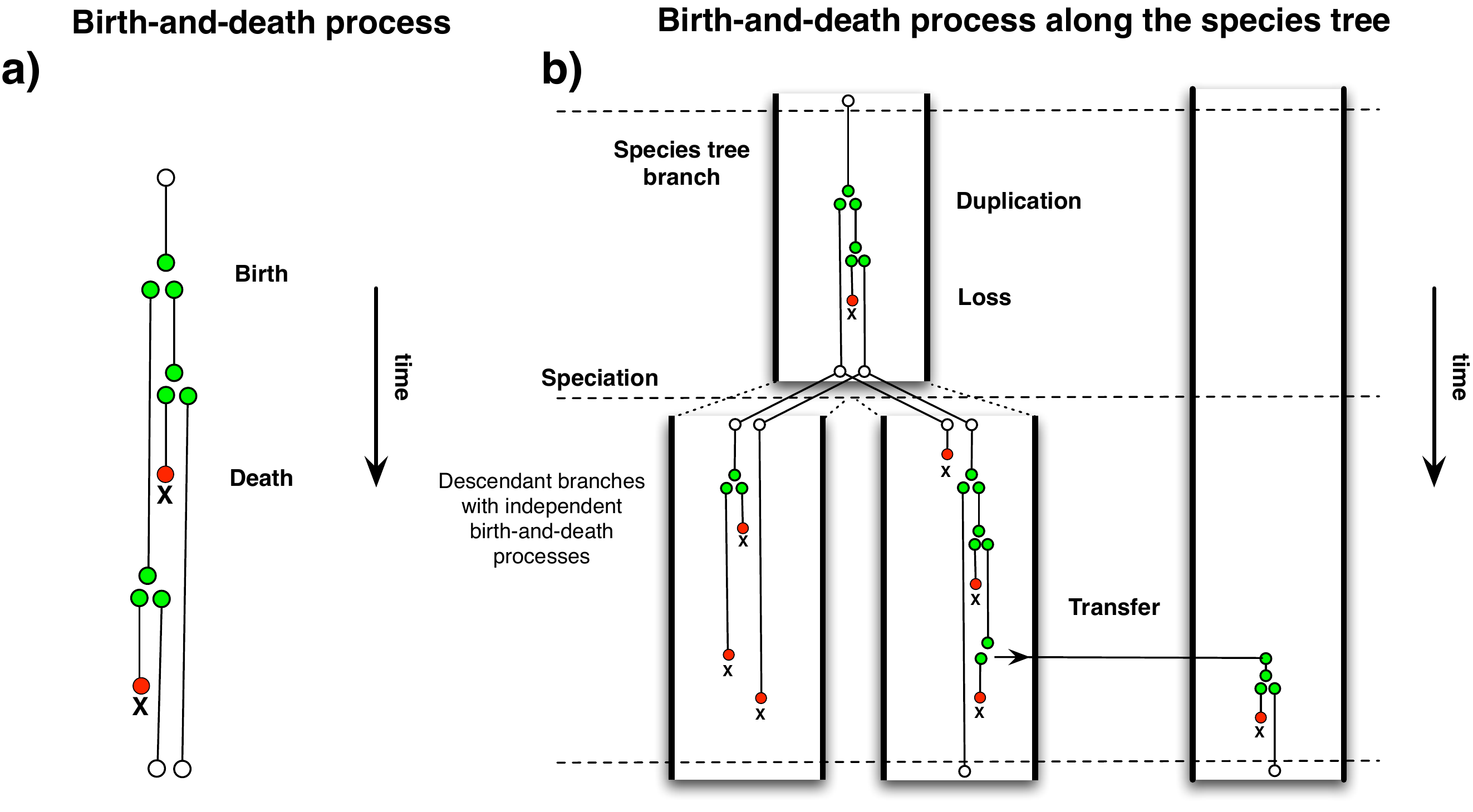}
		\caption{Birth-death processes for generating species trees and gene trees. Death events (species extinctions and gene loss) are in red, birth events (speciation, duplication and transfer events) are in green. a) Birth-death processes modeling speciation and extinction. b) Birth-death process modeling gene family evolution inside a species tree. }
		\label{fig:figure2}
	\end{figure}
\end{center}

Irrespective of whether they deal with ILS, DL, or T, all models of gene family evolution can be seen as generating a tree inside a tree, \textit{i.e.} a gene tree inside a species tree. In this respect, the models encountered in the literature dealing with processes of speciation and extinction \citep{Rannala1996,Morlon2009} are very similar to the models of gene family evolution. They both invoke birth-death processes that generate a rooted, clock-like tree topology. Birth events correspond to bifurcations, death events correspond to the loss of a lineage. In diversification models, lineages correspond to species, in models of gene family evolution they correspond to genes. After the birth-death process arrives at the present, lineages with no descendant among extant species are usually pruned, and the remaining lineages constitute the generated tree topology. Models of gene family evolution, however, are constrained by the species tree, whereas species diversification models are not. The species tree constitutes a set of constraints corresponding to speciation events and branch lengths that control the birth-death process generating the gene family tree. A gene family evolution model is in essence a series of species diversification models fitted piecewise along the branches of a species tree (Figure \ref{fig:figure2}). The birth-death process generating the gene tree starts above the root of the species tree. Each time the birth-death process reaches a speciation node, two new processes are created in the children lineages. These processes can be identical, or can have different parameters. In general, for $n$ branches in the species tree, counting the branch above the root, there are $n$ independent birth-death processes. Of course, the parameters of these $n$ independent processes do not need to be independent: one can imagine that the birth parameter for instance evolves according to e.g.\ a Brownian motion process running along the species tree. Such a model would penalize large jumps in the birth parameter between neighboring branches of the species tree, but to our knowledge such an idea has not yet been implemented. Another source of dependence between processes is lateral gene transfer: a birth in a lineage may originate in another.

The above describes how a gene tree, complete with branch lengths in units of time, is generated along the clock-like branches of the species tree. In practice, to simplify the problem, we will see that several methods choose to consider only the topologies of gene trees, \textit{i.e.} the branch length information is discarded. In this case, the mathematical machinery of the birth death model is not used to compute the probability of a specific dated scenario of observed birth and death events; instead it is used to compute the probability of a given succession of birth and death events \citep{Degnan2005,Wu2011}, or the probability of observing $k$ genes at the beginning of a branch of the species tree, and $l$ genes at the end of this same branch \citep{Boussau2013}. The choice of discarding branch length information in the gene tree simplifies the problem, because fewer processes need to be modeled. Potentially useful information, however, is discarded in the process.

%In the following, we review the models of gene family evolution that have been described in the literature.

\subsection*{The coalescent models population-level processes along the species tree.}
Coalescent models aimed at modelling the discordance between gene tree and species tree arising from a population-level process have enjoyed an increasing popularity in the last 10 years. Here birth events correspond to the appearance of a new allele, and death events to the disappearance of an allele, without any change in the locus of the gene. At any given time in a species, for a given locus in the genome, there may be several alleles. These alleles have their own history, some alleles being more closely related than others. When speciation occurs, most alleles will be sorted randomly between the two incipient species: in some cases both species will receive copies of all alleles, in others, each will receive only a subset of the alleles present in the parent population. In all cases, the history reconstructed from the allele sequences will be the allele history, and not the species history. The allele history and the species history always differ in the timing of the bifurcation events: assuming no hybridization has occurred between the lineages, alleles have necessarily split before species split. They can also differ in their topology, especially if only a brief interval of time passes between successive speciation events, and/or the effective population size of the parent species is large \citep{Rosenberg2002a}. Given a coalescent model, the amount of discordance in topology and divergence times between the trees of several loci and a species tree can therefore be used to estimate effective population sizes along the species tree \citep{Rannala2003,Liu2007,Heled2008,Kubatko2009,Minin2008}. Such a model where the population size is assumed to differ between the branches of the species tree has been called the multispecies coalescent \citep{Rannala2003}, and is now widely used to infer species trees given several loci, with one allele per locus \citep{Liu2007,Heled2008,Kubatko2009}, or several alleles per locus \citep{Liu2008,Heled2010}, using branch lengths information in the gene trees \citep{Liu2007,Heled2008}, or instead relying solely on their topologies \citep{Degnan2005,Wu2011}, in the maximum likelihood \citep{Kubatko2009,Wu2011} or Bayesian frameworks \citep{Liu2007,Heled2008,Liu2008,Heled2010,Minin2008}. The multi-species coalescent has also been used in conjunction with a Hidden Markov Model (HMM) running along a chromosome to infer recombination rates, locus-specific trees, divergence times and effective population sizes \citep{Hobolth2007,Mailund2011,Hobolth2011,Li2011,Scally2012}, given a species tree and a genomic alignment. In these models, gene trees can change along the alignment, and how often the gene trees change provides information on the rate of recombination.

One hidden assumption of all the models we have presented so far is that sequences are assigned to species without ambiguity. However, for recently diverged populations, it is often unclear whether two distinct populations should be considered as coming from the same single species, or should define two distinct species. In a given sample, it can also be unclear how many species have really been sampled. To address such cases, \citet{Yang2010} have developed a model for delimiting species given a set of sequences for different loci and several individuals. They assume a multi-species coalescent model, and postulate that hybridization does not occur after speciation. Information coming from \textit{e.g.} geographical distribution, morphology or behaviour can be introduced into the model thanks to a prior on the expected number of species, but also through a guide tree, whose other purpose is to improve the efficiency of the MCMC algorithm used for inference.  Under this model, they can analyse about 10 species for 100 sequences, with a finite number of loci. They apply this method to well studied data sets of asexual rotifers and fence lizards, and recover the species found by other means. %They also apply this method to human data from 6 populations, and as expected, their method infers that all individuals come from a single species.

\subsection*{Models of gene duplication and loss}	
Models of gene duplication and loss usually ignore the population-level processes (but see \cite{Rasmussen2012}, discussed below) that drive the fixation or disappearance of an allele, and only consider events of gene duplication and loss that have fixed in the species. In this setting, birth events correspond to fixed gene duplications, and death events to fixed gene losses. Probabilistic models for gene duplication and loss were first proposed by \citet{Arvestad2003}, and further developed in subsequent papers by the same group \citep{Arvestad2004,Akerborg2009,Sjostrand2012} and by a few others \citep{Dubb2005,Rasmussen2007,Rasmussen2010}. The focus of these works was to infer gene trees given a fixed species tree, with clock-like branch lengths in units of time, and with fixed rates of gene duplication and gene loss over the entire tree. Combined with the birth-death model of gene evolution is a hierarchical model of the rate of sequence evolution, wherein the species tree provides dates, and each gene family is associated with one or several rates. Alternatively, \citet{Gorecki2011,Gorecki2013} developed another model for gene duplication, not based on a birth-death process, but based on a Poisson process for computing the probability of a parsimonious reconciliation of a gene tree topology against a species tree. The gene tree does not need to have branch lengths, but the species tree does. For this reason, and because it does not include a loss parameter, this model misses some of the realism of the birth death processes described above, but gains in speed. 

More recently, we modified birth-death models to allow different duplication/loss parameters for each branch of a non-dated species tree \citep{Boussau2013}. To speed-up computations, we took an approach similar to \citet{Gorecki2011}, and did not account for the branch lengths of gene trees in their reconciliation with the species tree. Instead we only reconciled topologies. However, because our hierarchical model includes a model of sequence evolution for joint inference of gene trees and species tree, we still needed to estimate branch lengths in the gene trees. To simplify the problem, we chose not to have a hierarchical model of rates of sequence evolution: rates for each gene family were considered to be entirely independent. This decreased the number of global parameters to estimate, but increased drastically the number of gene family-specific parameters to estimate. 
	
	\subsection*{Models of lateral gene transfer}
Lateral gene transfer (LGT) corresponds to the incorporation in a genome of a gene coming from a different species. There is compelling evidence that this process has played an important role in the evolution of life, particularly in the domains of Bacteria and Archaea. Several models have been proposed to account for LGT in gene tree species tree reconciliation. So far they all consider events that have fixed and ignore population-level processes. One key feature of these models is whether they consider or not the possibility of gene replacement. The recipient of a transfer may, or may not have a gene homologous to the incoming gene in its genome. If it has, the gene in the recipient can be either conserved or lost. 
Transfer-only models usually consider only gene replacement. Models that do not make this assumption are often coupled with duplications and therefore are presented in the subsequent section. The difference between the two is important in that gene replacement is not modeled by birth and death. Indeed, replacing a lineage by a gene coming from another breaks the independence assumption between lineages, which is at the basis of computations in birth/death models. Thus models of gene replacement differ from the more mainstream birth-death models.

	The first attempt at modelling probabilistically lateral gene replacement explicitly was made by \citet{Suchard2005}, but a model for host-parasite cophylogeny developed by \citet{Huelsenbeck2000} could also be used to detect gene transfer between two loci. This model assumed that the parasite phylogeny differed from the host phylogeny through a Poisson-distributed number of host-switches, or replacements in our case. Transfer events could be placed uniformly among branches, or preferentially among branches close to each other in the host tree. Rates of evolution were independent in the host tree and the parasite tree, but sequence evolution was assumed to follow a strict clock model. Inference was conducted in a Bayesian framework through MCMC, and resulted in a host tree distribution, a parasite tree distribution, and distributions of the number and rates of host switches. 

	\citet{Suchard2005} proposed a model specifically designed to tackle gene replacement on a gene tree topology, discarding branch length information. Computing the probability of a gene tree given a species tree therefore did not involve mapping the gene tree onto the species tree. Instead, this method involved constructing several random walks between the gene tree topology and an unrooted species tree topology, each step of the walk corresponding to one Subtree Prune and Regraft (SPR) event \citep{Hordijk2005} (although another type of move was also considered). The probability of a gene tree given a species tree depended on the sum of the probabilities of the sampled paths, according to a Poisson model for the number of events. Using this model, \citet{Suchard2005} could estimate the species tree that best explained a forest of over 140 gene trees. However, the approach was limited to trees with only 6 to 8 species because of its computational cost. In addition, this approach does not make sure that all transfers detected in a set of gene families are time-consistent: in principle a gene can only be transferred to contemporaneous species, present in the sample analyzed or not, and certainly not to more ancient species. In \citet{Suchard2005}'s approach, because the species tree topology is not anchored in time, time-consistent transfers and "back-to-the-future" transfers are not distinguished. 
%In addition, SPR moves should only be used to model HGT on a rooted species tree because moves that include the root of the species tree are not biologically relevant [JE NE COMPRENDS PAS "biologically relevant"].

\citet{Bloomquist2010} chose another road to modeling gene replacement, by drawing on models of population genetics. They considered the Ancestral Recombination Graph (ARG). An ARG is a type of rooted network that combines both vertical and transfer edges. Once an ARG is built, it can be used to generate dated gene trees, which correspond to tree-like paths obtained by selecting edges of the ARG. They aimed at reconstructing an ARG that represents all of the evolutionary histories of a set of distinct loci, some of which evolved along the species tree, and some of which underwent replacement events. They used MCMC to propose ARGs, adding or subtracting transfer edges through reversible jumps. Given an ARG, a gene tree is drawn for each locus under study; these gene trees can be totally independent of each other, or can incorporate spatial information, so that two neighboring genes on the genome are more likely to be transferred together than distant genes. This allows modelling different situations, such as single-gene conversion or homologous recombination. In addition, the sequences of different genes can evolve at different rates. In the end, the method builds dated gene trees, and a dated ARG in which vertices are annotated as vertical or transfer nodes, and in which edges involved in a transfer event can be annotated with the genes that are inferred to have been transferred.

% More recently algorithmic progresses in the parsimonious reconciliation of a gene tree with a species tree in the presence of duplications, losses and transfers have been transferred to methods of statistical inference and can analyze several dozen species at a time (see below).

\begin{center}
	\begin{figure}[P]
	\centering
		\includegraphics[width=1.\columnwidth]{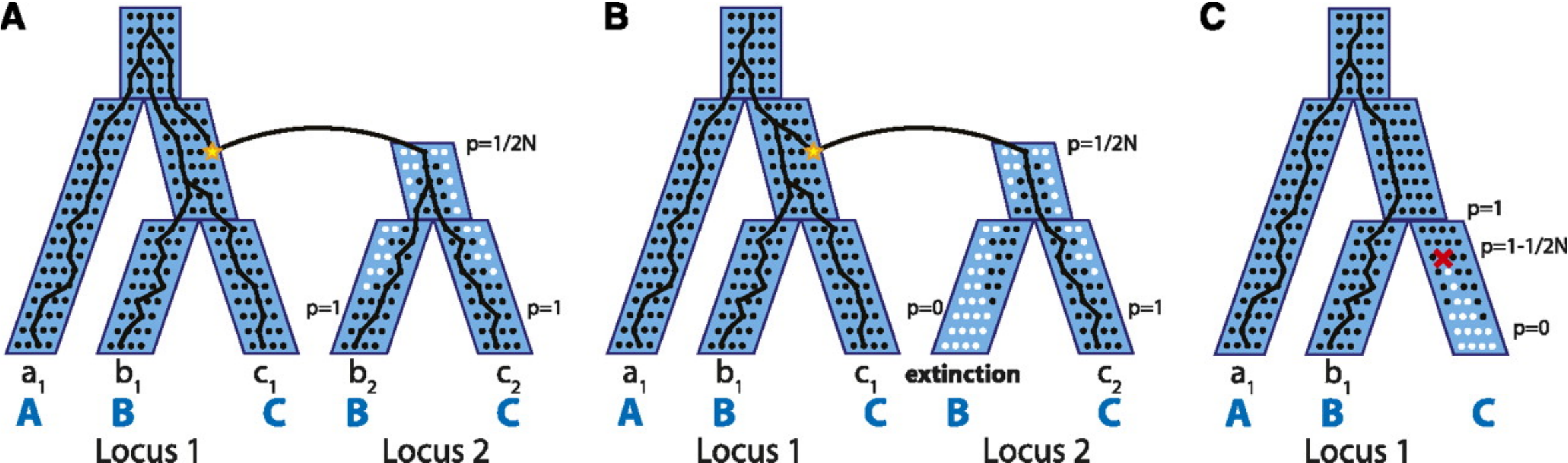}
		\caption{Duplication and loss events within a multispecies coalescent. (A) A duplication occurs in one chromosome and creates a new locus, locus 2 in the genome. At locus 2, the Wright-Fisher model dictates how the frequency p of the daughter duplicate (black dots) competes with the null allele (white dots) until it eventually fixes (p = 1). A gene tree is therefore a traceback in this combined process. (B) A new duplicate can undergo hemiplasy, and fixes in some lineages and goes extinct in others. (C) Similar to duplication, a gene loss (deletion) starts in one chromosome and drifts until it fixes or goes extinct. Reproduced from Figure 1 in Rasmussen M D , and Kellis M Genome Res. 2012;22:755-765.}
		\label{fig:locustree}
	\end{figure}
\end{center}

	\subsection*{Models that combine the above}
We know of five probabilistic models that combine some of the processes listed above: the DTLSR model of \citet{tofigh_2009}, the DLCoal model of \citet{Rasmussen2012}, our ODT model \citep{Szollosi2012,Szollosi2013}, network-based models of hybridization with incomplete lineage sorting \citet{Than2007,Meng2009,Kubatko2009,Yu2012}, and finally network-based models of polyploidization \citep{Jones2013}. 

The DLCoal model combines a coalescent model with a model of gene duplication and loss. In doing so, it reintroduces population genetics concepts into the framework of DL models: for instance, it acknowledges that a new duplicate first has a very low frequency in a population, as it is present only in the individual where the gene duplication occurred. Under this model, the reconciliation of a gene tree with a species tree requires three objects: first, a dated species tree, with branches annotated with effective population sizes. Second, a dated locus tree, generated thanks to a birth-death process placing events of duplications and losses along the species tree branches, according to a duplication rate and a loss rate. The locus tree contains implicit information about chromosomal positions, as under this model chromosomal position changes at duplication nodes, but is generated according to the same birth-death process used in previous duplication-loss models \citep{Arvestad2003,Arvestad2004,Akerborg2009,Sjostrand2012,Dubb2005,Rasmussen2007,Rasmussen2010}. Third, the gene tree, generated according to a coalescent process running along the locus tree (see Fig.\ref{fig:locustree}). The DLCoal model makes the simplifying assumption, termed the ``hemiplasy assumption'', wherein all duplications and losses are considered to either always go extinct or never go extinct in all descendant lineages. This assumption allows the separation of the duplication-loss process from the multilocus coalescent. Aside of this assumption the model entails two successive tree-embedded birth-death processes and can be regarded as a multispecies colaescent process taking place along the locus tree which in turn is generated along the species tree by a DL process (cf.\ Fig.\ref{fig:locustree}). \citet{Rasmussen2012} implemented this model into a program that can reconcile a gene tree with a species tree, given rates of duplication and loss, a dated species tree, and effective population sizes. 

The DTLSR \citep{tofigh_2009,Tofigh2011,Sjostrand_2013} and ODT models combine a model of gene transfer with a model of gene duplication and loss. Both are natural extensions of the DL models, and do not require an additional object like the locus tree in the DLCoal model. Instead, the birth-death process is modified to accommodate two types of birth events, gene duplications and gene transfers. 
%The DTLSR model of \citet{Tofigh2011} has to date only been published as part of a PHD thesis with limited validation and application to biological data. 
The ODT model has been applied to biological datasets, and has been extended to account for gene transfers that involve species that have gone extinct or are otherwise unrepresented in the tree under consideration \citep{Szollosi2013}. In this extension of the ODT model, the ex-ODT model, sampled species and their ancestors are assumed to come from a population of species, which at any given time contains many more species than are present in the sample. This population of species evolves according to the Moran process, but could evolve according to more complex processes (e.g. \citep{Morlon2009}). Within this framework, a gene can be transferred from the genome of an ancestor of a sampled species to the genome of a species living at the same time, but which gave no descendant among the sampled species. Then this gene can evolve among the genomes of species that gave no sampled descendant, including through transfers, duplications, losses and speciations, and finally reintegrate genomes with sampled descendants. 

\begin{center}
	\begin{figure}[P]
	\centering
		\includegraphics[width=1.\columnwidth]{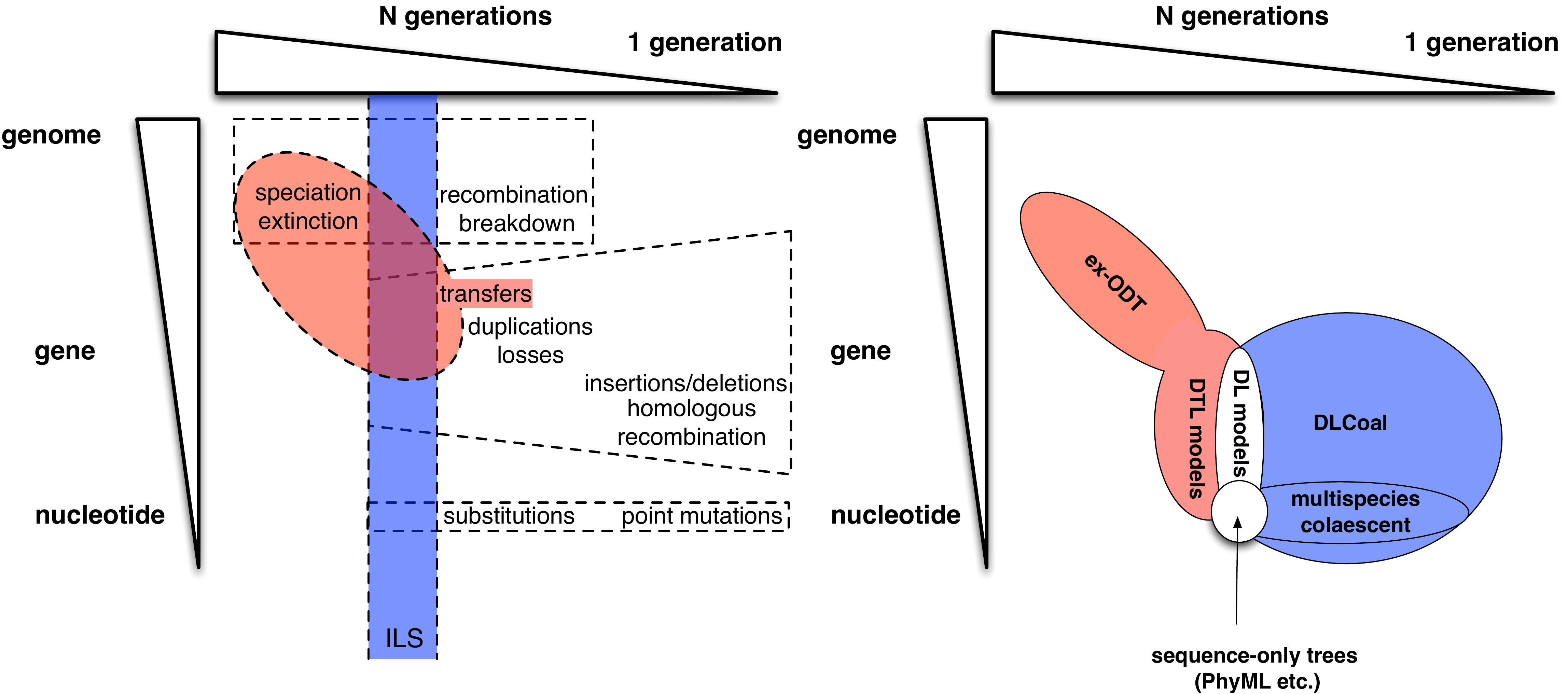}
		\caption{ Left: The hierarchy of sequence evolutionary processes plotted along two dimensions: sequence length, from single nucleotides thru genes to whole genomes; and time, from single generations thru the neutral fixation time to large numbers of generations. Events occurring in single individuals, such as point mutations, insertions/deletions etc.\ are filtered by the population level process of selection and drift with only a minority reaching eventual fixation. Speciation and extinction events affect entire genomes and require many generations. Incomplete lineage sorting (blue) occurs when fixation time overlaps with speciation time. Transfer events can cross large phylogenetic distances and almost always involve evolution along extinct or unrepresented species and hence are affected by speciation dynamics.  Right: Distribution of models discussed in the text. Classic molecular phylogeny methods model only substitutions. DL models handle fixed duplication and loss events along with substitutions. The multispecies coalescent and related methods model explicitly the fixation of point mutations (blue). DLcoal models the fixation of both point mutations and of gene scale of insertion/deletion events that lead to fixed duplication and loss events. DTL models (red; ODT, DTLSR) extend DL models to fixed transfer events, but ignore speciation dynamics. The ex-ODT model combines speciation dynamics with DTL events to provide a more realistic model of transfer paths. A potential "ex-DTLCoal" model, as discussed in the main text would cover the area of all these models.}
		\label{fig:models}
	\end{figure}
\end{center}

Contrary to the DLCoal model, neither the ODT models or the DTLSR model account for population-level processes, and thus do not model allele fixation. However, the two model types could be easily mixed, e.g.\ the ex-ODT model could be used to generate a locus tree, which would then be used to generate a gene tree according to the coalescent model. This ex-DTLCoal model would then account for speciations, extinctions, duplications, losses, transfers and incomplete lineage sorting in a hierarchical probabilistic model (cf. Fig.\ref{fig:models}).

\citet{Meng2009,Kubatko2009,Yu2012} propose a model for the detection of hybridization in the presence of incomplete lineage sorting, extending early efforts by \citet{Than2007}. As we noted above, from a modeling point of view, hybridization may be seen as a high frequency of gene replacements between two lineages. There, a rooted phylogenetic network is used instead of a phylogenetic tree. In a hybrid genome, any gene is assumed to be coming from one of two parental genomes. Hybridization nodes are thus represented in such a network by nodes with two parents $A$ and $B$. A probability $\gamma$ indicates the proportion of genes coming from parent $A$, the rest coming from parent $B$.  Under this model, evolution of alleles within the network is very similar to their evolution along a species tree. At nodes with a single parent, the usual multi-species coalescent model applies, with parameter the length of the branch, in coalescent units. At hybridization nodes, parent $A$ is chosen with probability $\gamma$, otherwise parent $B$ is chosen; then the usual multi-species coalescent applies with the chosen parent. \citet{Yu2012}, building upon \citet{Meng2009} who worked with a single hybridization event, provide formulas for computing the likelihood of a gene tree topology (without branch lengths) under this model, for networks with any number of hybridization nodes, which makes it possible in principle to carry out gene tree inference or species tree inference. In practice, they implemented this model in PhyloNet, so that different candidate species trees or networks can be compared according to their likelihood given a set of gene tree topologies, or a set of gene tree topology distributions. \citet{Kubatko2009} implemented an algorithm to search for the optimal network according to information criteria, with a model that considers branch lengths in the gene trees in addition to their topologies. In this implementation, the topology of the network is fixed, and the object of inference is the presence and number of hybridization nodes with the associated parameters $\gamma$.

Allopolyploidization has also been modeled as a specific case of hybridization \citep{Jones2013}. In this context, allopolyploidization occurs when two diploid individuals from different species mate, which results in the birth of a viable new tetraploid species. \citet{Jones2013} make the simplifying assumption that there is no recombination between the alleles inside the allopolyploid, and propose two models. In one model, any number of allopolyploidization events can be inferred, but evolution in the two genomes forming the allopolyploid is assumed to be independent, which disregards the fact that these genomes belong to the same species. In the second model, only one event of allopolyploidization can be inferred, but then the evolution of the two genomes in the allopolyploid is linked. In both cases they use the multispecies coalescent model to describe the evolution of alleles along the species phylogeny or, in this case, the species phylogenetic network.They apply these models to simulated data sets as well as empirical data sets of less than 10 species, less than 10 genes, and up to three alleles per gene.

\subsection*{Simulation and inference}

\begin{center}
	\begin{figure}[P]
	\centering
		\includegraphics[width=1.\columnwidth]{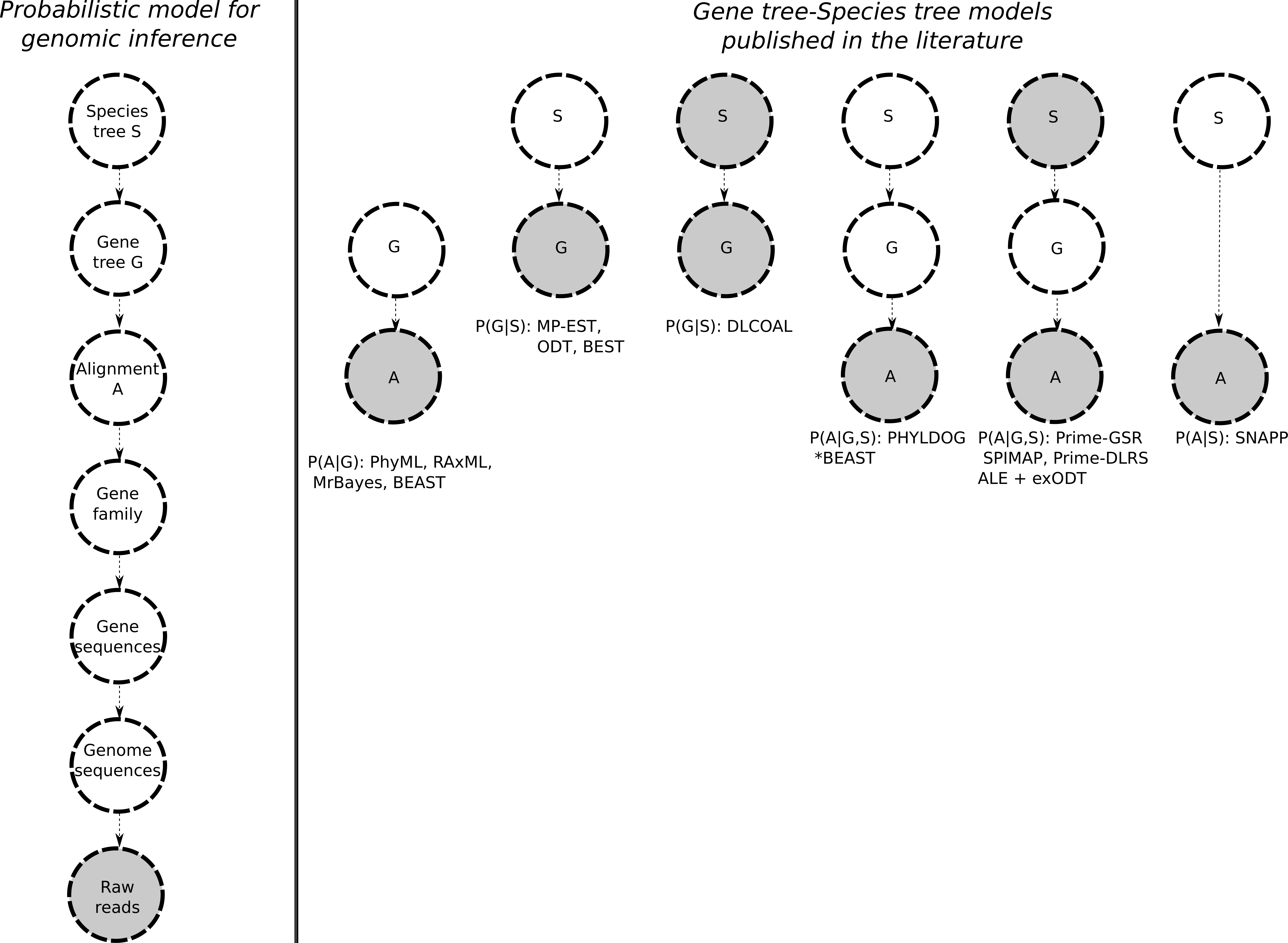}
		\caption{Gene tree-species tree models in the context of the phylogenomics inference pipeline. Left: the inference pipeline  (some steps are not represented, such as sequencing error correction). Right:  graphical representation of the inferential problem for a selection of the models and associated phylogenetic software discussed in the main text. The likelihood that must be computed is also shown. Graphical model conventions are observed: stochastic nodes, coming from probability distributions, are represented with dashed lines, nodes corresponding to data considered as known are grey, and nodes whose states are inferred are in white. The models have been simplified, and parameters others than the gene tree and the species tree have not been represented.}
		\label{fig:graphModels}
	\end{figure}
\end{center}

	One can see a phylogenetic pipeline as a series of statistical inferences, starting from raw sequences coming out of sequencing machines, and finishing with the inference of a species tree (Figure \ref{fig:graphModels}). Necessary steps include sequencing error correction, assembly of reads into contigs and scaffolds, gene annotation, gene family clustering, alignment, and tree reconstruction. Most of these steps are done sequentially, so that late steps in the pipeline entirely disregard any estimate of uncertainty from the previous steps, and do not provide any feedback to these. Gene tree-species tree models take a step towards a more principled approach, by allowing communication between two steps of this pipeline, the construction of gene trees, and the construction of a species tree. Figure \ref{fig:graphModels} places the above discussed models and associated phylogenetic software in the context of the complete phylogenetic inference pipeline. Grey nodes are considered known, and white nodes are inferred. This figure shows that a large diversity of inferential problems have been adressed, considering gene alignments, gene trees, species trees, or several of these as data. However, the first steps of the pipeline are often made without a solid statistical framework, in complete isolation from each other. In this section we review some of the methods and algorithms that have been used to address these inferential problems. We present how data can be simulated, how the likelihood of a gene tree or of a species tree can be computed efficiently, and how good gene trees and species trees can be searched for.

\subsubsection*{Simulating gene trees}
%Here, we detail a possible implementation for a diversification process, the multispecies generalizations of which involves traversing a species tree from its root to its leaves and running the processes on each branch.

Generating a simulated gene tree according to a model is straightforward, and involves traversing a species tree from its root to its leaves. Assume there are $k$ lineages at time $t_{begin}$ (which at first is the beginning of a branch) on a branch $i$ which ends at $t_{end}$, with birth rate $\lambda_i$ and death rate $\mu_i$. Two waiting times $t_{birth}$ and $t_{death}$ are randomly drawn from exponential distributions, one with parameter $k*\lambda_i$, and the other with parameter $k*\mu_i$. If both are longer than $t_{end}-t_{begin}$, no event occurs along this branch. If $t_{birth}$ is the shortest, then a birth event occurs, and $t_{begin} \leftarrow t_{birth}$ and $k \leftarrow k+1$. If $t_{death}$ is the shortest, then a death event occurs, and $t_{begin} \leftarrow t_{death}$ and $k \leftarrow k-1$. In both cases, the lineage that has undergone the event is chosen uniformly. Then the process starts again. Lineages generated at the end of a branch of the species tree are then given as input to the next birth-death processes, running along the descendant branches.

Lateral gene transfer can be regarded as simply a peculiar birth event, one which results in the birth of a gene copy in a  branch of a species tree different from the species tree branch of the ancestral copy. Computationally this introduces a dependency between species tree branches, requiring that all contemporaneous branches are considered together. To simulate gene trees one can then consider different rates for different birth events, \textit{i.e.} for duplications and transfers \citet{Szollosi2012}. Alternatively, one can consider replacement transfer, wherein, if a member of the homologous gene family is present in the recipient genome, it is replaced by the transferred gene.  Computationally this introduces a dependency between gene tree branches that prevents the use of algorithms that rely on the independence of gene lineages (see below), but simulations can be carried out in a straight-forward manner \citep{Galtier2007b}. More problematically, however, no simulation method has been constructed to take into consideration the fact that, in the presence of transfer, gene trees record evolutionary paths along the complete species tree, including extinct and unsampled branches, and not only along the phylogeny of the species in which their descendants reside today. This is the case because, as first noted by \citep{Maddison1997} and later elaborated by \citet{Zhaxybayeva:2004fk,Fournier:2009nx}, while transfer events imply that the donor and receiver lineages existed at the same time, the donor lineage might have subsequently become extinct, or more generally, might not have been sampled. Brute force simulation of transfers along a "complete phylogeny" would be very expensive due to the large number of species that must be considered. There are at least two possible alternatives: i) it is possible to use instead parametric bootstrap like methods described below or ii) approximations based on the assumption that the number of species represented in the species tree is much smaller that the total number of species \citet{Szollosi2013} could potentially be exploited similar to the coalescent.   

A parametric bootstrap-like approach was used by  \citep{Szollosi2013} in the context of the ALE+ex-ODT model to produce simulated alignments based on a species tree and real alignments from 36 cyanobacteria. The approach consisted of reconstructing the most probable gene trees and associated duplication, transfer and loss rates given a fixed species tree and the gene family alignments (for discussion of how this conditional probability is computed see below). The inferred gene trees were then used to simulate alignments. These were used to assess the reconstruction accuracy of the ALE+ex-ODT model, comparing the reconstructed gene trees and associated duplication, transfer and loss events given the fixed species tree and the simulated alignments to those used in the simulation. This approach has the advantage of circumventing potentially complex simulations while at the same time retaining otherwise hard to reproduce properties of biological datasets, such as the distribution of gene family sizes and the variation of evolutionary rates within and among gene families \citet{Szollosi:2012uq}.       

\subsubsection*{Computing the conditional probability of a gene tree}

By the joint conditional probability of a gene tree, we mean the probability of a gene tree given a gene alignment and a species tree. There are (at least) two components to the conditional probability of a gene tree. One component corresponds to the model of sequence evolution running along the gene tree; the other to the model of gene family evolution running along the species tree. In both cases, dynamic programming algorithms traversing the nodes of gene trees and species trees can often efficiently compute the relevant component of the probability.

Along a branch of a tree of a given length (the gene tree for sequence based models or the species trees for gene family evolution models), probabilities of descendant states given ancestral states are computed by solving differential equations similar to other birth-death processes (for some processes, the solution can be obtained analytically, in others, numerical integration is necessary). Dynamic programming is then used to traverse branches of the tree in postorder. That is, branches are considered starting from their leaves up to the root. If a tree is unrooted, nodes need to be visited three times (although only two tree traversals are necessary, \citep{Guindon2003}), according to the three possible directions for the root.

The probability of an alignment given a gene tree is computed along the gene tree alone, while the probability of a gene tree given a species tree is computed along both the gene tree and the species tree. In both cases, at a leaf, data can be used to fill a vector of probabilities. For sequence evolution, data corresponds to the state found at the site under consideration (\textit{e.g.} A, C, G or T in a DNA sequence). For the model of gene family evolution, this corresponds to the presence, absence or number of genes found in a given extant species. Then, at internal nodes, probability distribution vectors from the children nodes are used to compute the probability of a given subtree, according to the process considered in the branches. At the root, dynamic programming algorithms yield a probability for the entire tree. 

The rough description above outlines the algorithm developed by \citet{Felsenstein1981} to compute the probability of a multiple alignment of gene sequences given a gene tree and a model of sequence evolution. In this case the differential equations corresponding to the Markovian process of sequence evolution can be solved analytically to obtain substitution probabilities along a branch of a given length. Computing the probability of a gene tree given a species tree is a bit more complicated, as it involves mapping the gene tree onto the species tree to compute the probability of presence of a gene tree node or branch at each node or branch of the species tree.  This mapping is natural at the leaves: a gene from species $A$ is mapped onto leaf $A$ of the species tree. For internal nodes, the mapping can be helped by the consideration of node ages in models that consider that both the gene trees and the species tree are dated.  This is typically the case with multi-species coalescent models (\textit{e.g.} \citep{Rannala2003}). Such a method yields a single mapping between the nodes of a given gene tree and a given species tree, for given rates of sequence evolution. In a duplication and loss context, \citet{Akerborg2009} improved upon this approach by analytically integrating over the possible mappings as well as over rates of sequence evolution, again through dynamic programming. Their approach requires "slicing" the species tree by dropping extra nodes along the branches of the tree. These two methods yielding either a single mapping or integrating over all mappings in the context of dated trees have counterparts in the context of non-dated trees. On one hand, \citet{Boussau2013} assumed the most parsimonious mapping between the nodes of the gene tree and the nodes of the species tree. This most parsimonious mapping is obtained with a single tree traversal \citep{Zmasek2001}. On the other hand, \citet{Szollosi2012} took a similar approach to \citet{Akerborg2009} by integrating over all the possible mappings between the nodes of the gene tree and the nodes of the species tree, again through dynamic programming, but without considering dated gene trees. This allowed them to avoid using a model of rate evolution. However it was necessary to order the nodes of the species tree, which has the effect of slicing it and adding new nodes, for correctly computing the probability of a gene tree given a species tree. As this inference includes transfer aside of duplication and loss, numerical integration is necessary to solve the differential equations describing the birth and death process because gene lineages mapping to different branches of the species tree are dependent and no analytical solutions are available.

Usually such algorithms can achive linear complexity in the number of genes for coalescent or DL models, while handling transfers raises the complexity to the product of the number of nodes in the two trees. Methods that require slicing the species tree \citep{Akerborg2009,tofigh_2009,Doyon2010,Szollosi2012} introduce new nodes and therefore are more expensive. In models with transfers, slicing the species tree is nonetheless necessary as computing the probability of a gene tree given a species tree is provably difficult when the species tree nodes are not ordered, except if time-inconsistent transfers are allowed \citep{Tofigh2011}. Dynamic programming also fails when transfers replace existing lineages (like in SPR-like events of \cite{Suchard2005,Nakhleh2005,Beiko2006,Bloomquist2010,Abby2012}) because this introduces a dependency between gene tree lineages. In this context the space of gene trees must be explored using SPR-like moves to compute the probability of a gene tree given a species tree. Note that methods that slice the species tree \citep{Tofigh2011,Szollosi2012,Szollosi2013} use a workaround to handle gene replacement transfers: they assume that a transfer event followed by a loss in the receiving lineage occurred.

For models of sequence evolution as well as models of gene family evolution, the same dynamic programming scheme can be used to make a variety of inferences. If we focus on models of gene family evolution, this scheme can be used to obtain a maximum parsimony reconciliation or the reconciliation that maximizes the probability of the gene tree given the species tree, to integrate over all reconciliations to compute the probability of a gene tree given a species tree, or to sample among reconciliations according to their probability. For a survey on available algorithms and software for reconciliations as of 2 years ago see \cite{Doyon2011}.

Combining the probability of a gene alignment given a gene tree with the probability of the gene tree given a species tree can be achieved by the multiplication of the two probabilities \citet{Maddison1997,Akerborg2009,Szollosi2013}, assuming that the gene alignment is independent of the species tree conditional on the gene tree. The same assumption is at the heart of the model by  \citet{Rasmussen2012} who combine probabilities from a multispecies coalescent model and a DL model, through the addition of an additional layer, the locus tree (see section "Modeling the dependence between gene tree and species tree"). Thanks to two conditional independence assumptions, the probability of the entire structure is obtained by the product of three probabilities. 
%Including transfer in this framework is the subject of prospective works \cite{Stolzer2012}.

Computing the probability of a gene tree given a phylogenetic network according to a multi species coalescent model, as in \citet{Yu2012}, requires specific algorithms in addition to usual ones, and the introduction of a new type of tree, the MUL tree. The MUL tree is a multi-labeled tree generated from a phylogenetic network as follows. Every hybridization node with its two parents $A$ and $B$ is removed from the tree, then duplicated, and finally one copy is attached to $A$ and the other to $B$. The MUL tree therefore contains several copies of subtrees coming from hybridization nodes, but it is no longer a network, as all nodes have a single parent. The computation of the probability of a gene tree then involves usual dynamic programming algorithms running along the MUL tree, with two complications. One is that the multi-species coalescent process must be aware of the number of alleles evolving in a duplicated subtree when computing the propagation probability of an allele; the second is that the hybridization probabilities $\gamma$ must be taken into account (see the "Modeling the dependence between gene tree and species tree" section for more details about $\gamma$) to compute the probability of an allele trajectory along the MUL tree. MUL trees are also used by \citet{Jones2013}, as an intermediate step to compute the probability of a phylogenetic network in their most accurate model that can handle a single allopolyploidization event, and as object of inference in their more flexible but less faithful model.

\subsubsection*{Finding good gene trees}

If we don't assume gene trees to be known, exploring the space of possible gene trees is usually achieved, similar to sequence evolution models, by either hill-climbing maximum likelihood strategies \citep{Vilella2008,Thomas2010,Rasmussen2012,Boussau2013,Wu2013}, or stochastic sampling of trees using Markov Chain Monte Carlo (MCMC) algorithms \citep{Liu2007,Heled2008,Heled2010,Minin2008,Akerborg2009,Rasmussen2010}.

These local searches are inspired by classic gene tree search algorithms \citep{Guindon2003}, and the MCMC or hill climbing steps use, for example, random NNI (nearest neighbour interchanges) or SPR (subtree prune and regraft) moves. As such searches can be computationally intensive, SPRs are sometimes bounded \citep{Boussau2013}, or rearrangements are directed \citep{Szollosi2012}. Devising good directed rearrangements is sometimes called gene tree correction, and entails for example decreasing the number of duplications in a duplication/loss most parsimonious reconciliation: such moves have some chance to increase the probability of a gene tree according to the model of gene family evolution. If at the same time they do not decrease the likelihood according to the model of sequence evolution by a large portion, such moves are accepted \citep{CHANG-EULENSTEIN06,Muffato2010,Chaudhary2012,Lafond2013}. Reconciliations with polytomies in a gene tree \citep{Lafond2012} can also be used to correct or construct good gene trees according to a sequence model and a species tree.

\begin{center}
	\begin{figure}[P]
	\centering
		\includegraphics[width=1.\columnwidth]{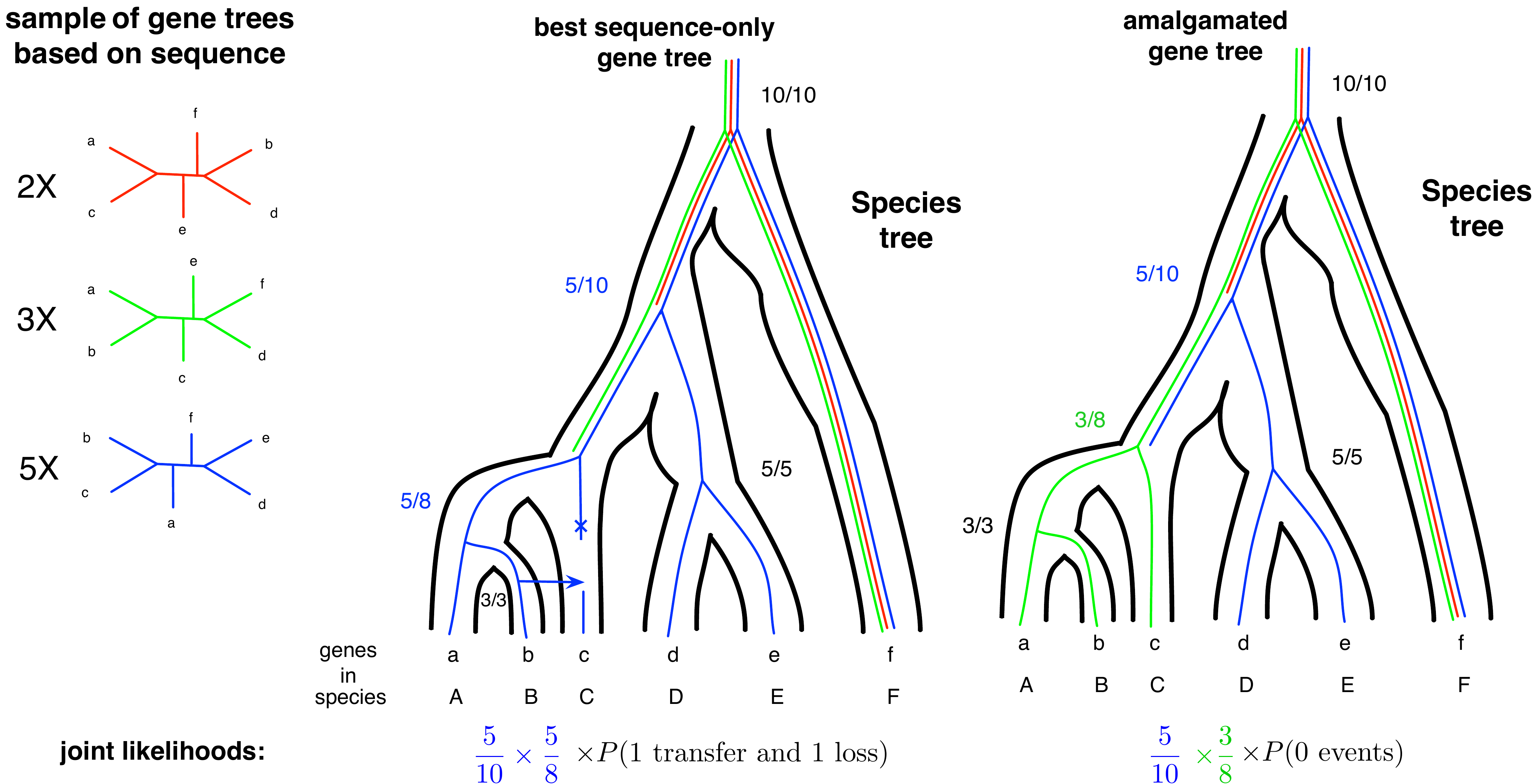}
		\caption{ Based on gene trees (red, green and blue trees) sampled according to their posterior probability, conditional clade probabilities (CCP) can used to estimate the posterior probability of any tree that can be \emph{amalgamated} \citep{Hohna2012}  from clades present in the sample. Conditional clade frequencies can be used to approximate CCPs and are computed as the fraction of times a particular split of a clade, e.g.\ (abc,de) is observed among all trees in which the containing clade, e.g.\ (abcde) is found. Estimates based on the sample of trees on the left are shown as fractions for two different gene trees that can be amalgamated. The estimate for a gene tree is given by the product of the frequencies. Amalgamated likelihood estimation (ALE, \citep{Szollosi:2013fk}) is a probabilistic approach to exhaustively explore all reconciled gene trees that can be amalgamated as a combination of clades observed in a sample of gene trees. Based on the sample on the left the tree with the highest posterior probability is the blue tree. Reconciling it with the species tree requires $1$ transfer and $1$ loss event. It is, however, possible to combine clades present in the blue and green trees to produce a gene tree that is not present in the original sample but is identical to the species tree, \textit{i.e.}\ it requires $0$ events to draw it into the species tree. Depending on the relative probabilities of $P(0~\mathrm{events})$ and $P(1~\mathrm{transfer}~ \mathrm{and}~1~\mathrm{loss})$, the joint conditional probability may prefer the scenario without transfer.}
		\label{fig:amalgamation}
	\end{figure}
\end{center}

An alternative to local search in gene tree space is the amalgamation of reconciled gene trees from a sample of trees \citep{David2011,Szollosi:2013fk}. As illustrated in Fig.\ref{fig:amalgamation} this approach consists in combining clades found in a sample of gene trees based on a model of sequence evolution (e.g. sampled using MCMC or by some resampling strategy such as bootstrap) in order to find the optimal gene tree according to both the model of sequence evolution and the model of gene family evolution. The probabilistic application of this method relies on the observation \citep{Hohna2012,Larget2013,Szollosi:2013fk} that it is possible to efficiently and accurately approximate the probability of an alignment for a very large number of gene trees using conditional clade frequencies based on a much smaller sample of trees. Combined with a model of gene family evolution, this allows the construction of a gene tree that maximizes the product of the probability of the alignment given the gene tree and the probability of the gene tree given the species tree, or alternatively to sample reconciled gene trees according to the joint conditional probability \citep{Szollosi:2013fk}. Here also a dynamic programming algorithm is used and the computation of a gene tree that maximizes the joint conditional probaiblity is polynomial in the number of trees in the sample, which can be much faster than a local exploration, and can also be used as a starting point for local exploration.

\subsubsection*{Finding a good species tree}

One can also assume that the species tree is unknown, and search for it. Indeed integrative models of evolution should be able to retrieve the information about species evolution from the alignment better than averaging methods like concatenations or supertrees.

If we assume idependence between genes, much like we usually assume independence between sites in most models of sequence evolution, a score for a species tree can be computed by adding (in parsimony) or multiplying (in probabilistic contexts) the joint conditional probabilities for all gene trees. Optimization algorithms can then be used to search for the species tree with the best score.

Methods for species tree inference under coalescent models (minimizing the number of ILS events or
maximizing a joint joint conditional probability for a set of gene trees) are reviewed by \cite{Liu2009b}
and can be found on the online ressource \cite{Shaw2013}. Fast approximations
are given by distance or supertree methods \citep{Than2009,Liu2010a,Liu2011,Yu2013},
while Bayesian sampling is
more precise but computationally intensive \citep{Heled2010,Liu2008,Kubatko2009}.

In a duplication and loss framework, \citet{Wehe2008, Bansal2010,Chaudhary2010} use the total number of duplications
and losses as a global score and propose an efficient way to perform SPR (subtree prune and regraft) and tree bisection reconnection (TBR) moves on one candidate species tree to decrease
the score. By performing SPRs in a specific order, they show that only a small portion of the mapping between gene trees and species trees needs to be recomputed, hence resulting in significant savings in computing time.

Alternatives to local searches are the search for exact solutions \cite{Chang2011}, or, inspired
by coalescent models, super-tree methods resembling the amalgamation of gene trees mentioned in the
previous section \cite{Bayzid2013}, which seem to quickly provide good approximations of parsimonious
species trees.

Considering transfers in a probabilistic framework, \citet{Szollosi2012} explored time ordered species trees, that is, species trees in which internal nodes are totally ordered. Topology search was performed by a directed local search guided by apparent highways of transfers: rearrangements are proposed in parts of the species tree that show the highest numbers of transfers in the hope of proposing rearrangements reducing phylogenetic discord. %This shows the possibility to infer, in addition to the species tree, a chronology of speciation events.

All of the above methods take as their input fixed gene trees. But we have several times recalled that good gene trees are only computed with the help of a good species tree. Joint estimation in a duplication and loss model has been achieved by \citet{Heled2010,Boussau2013}, but with a very high computational cost. Improvements in the algorithms used would be very welcome.

\section*{Impact on systematics and genome evolution}

The methods we described in the previous sections have shown repeatedly that they improve on methods that do not model gene family evolution, for species tree estimation, gene tree estimation, and the study of genomic evolution. In this section we present some of their most salient results.

\subsection*{Learning about species relationships and history}
Coalescent models have been used extensively to investigate species trees (e.g. \citep{Gray2011,Alstrom2011,Reid2012,Rocha2013}), because contrary to supertree or concatenation methods they should be robust against incomplete lineage sorting effects \citep{Degnan2006}. However, methods that jointly infer gene and species trees remain limited in their ability to handle large data sets: they cannot handle more than a few dozen species, and a few dozen loci. In their place, approximate methods have had to be used on genomic-scale data sets. As these take gene trees as input, they require much less computer resources, but suffer from the propagation of errors made during gene tree inference. \citet{Song2012} used MP-EST to reconstruct a well-resolved mammalian species tree based on 447 genes from 37 species. They found that the traditional technique, which consists of concatenating the alignments for several loci, was significantly less consistent when run on subsets of the data than MP-EST. However, another study by \citet{Mccormack2013} found a highly unresolved tree when they used the same program on a data set of 416 Ultra Conserved Elements (UCEs) form 32 species of birds. They assumed that the small size of the UCEs led to unresolved and erroneous gene trees, which in turn caused MP-EST to estimate a highly unresolved species tree. This illustrates that methods that do not infer gene trees, but rather obtain them from external sources are by construction very sensitive to the quality of the input gene trees, and calls for more accurate and robust methods able to jointly infer gene and species trees in the coalescent framework, for large data sets.

Models of gene duplication and loss, or of gene transfer, have also been used to reconstruct species trees, although more sparingly than coalescent-based models. The construction of a species tree given many fixed gene trees has been performed many times under a parsimony DL framework, searching for the species tree that minimises the total number of duplications or duplications and losses. In that context phylogenies have been proposed for many clades \citep{Slowinski1997,Page2002,Cotton2003,Than2008}. Genome wide scale was reached by \cite{Burleigh2011} who propose a plant phylogeny with 18,896 gene trees. So far, these models have mostly provided species phylogenies that were consistent with the literature \citep{Suchard2005,Szollosi2012,Boussau2013}. However, perhaps one of the most surprising benefits of modeling gene family evolution comes from modeling lateral gene transfer. Gene transfer is often described as a mere nuisance, that prevents phylogeneticists from obtaining well-resolved and easy to interpret species trees. According to this viewpoint, modeling gene transfer is useful because it provides a principled way to discriminate between vertical descent and lateral transfer: lateral transfers can then be discarded to focus on vertical descent and obtain a species phylogeny. However, gene transfer also provides additional information about ancestral species and their history.
	
  For example \cite{David2011} infer the gene birth rate along a deep phylogeny of prokaryotes, and conclude that 25\% of the genes in their data set were born in the Archean. They used a dated tree to infer transfers. But transfers can help date species trees, because gene transfers can only occur among contemporaneous species, and then be inherited by descendant species. A pattern of gene transfers therefore orients a species tree, from ancestral species that gave genes but did not receive many, to more recent ancestors that received genes from more ancient species.

	Based on this idea, we ordered speciation events in Cyanobacteria using 8,332 genes in 36 species \citep{Szollosi2012}. We found that the information provided by transfer events supported a root different from the root obtained using outgroup species. However, outgroup sequences are usually very distant from Cyanobacteria, and the choice of the outgroup species affects the rooting of the tree. In addition, we find that support for our unusual root comes from more than 200 transfer events. Overall the information gained thanks to the use of a model of gene family evolution provides a new light into the order of speciation events in Cyanobacteria. It also provides a unique insight into genomic evolution in this clade, by providing an accurate reconstruction of ancestral gene contents. Because the ODT model infers events of gene transfers, duplications and losses, the number of genes present in ancestral genomes in each gene family is a natural outcome. Future analyses of ancestral gene contents based on models like ODT should provide windows into ancient metabolisms and phenotypes.

%\subsection*{Detection of hybrid speciation with phylogenetic networks}
Another important process shaping species relationships is hybridization. Models that aim at inferring hybridization in the presence of incomplete lineage sorting have been used in several systems and have often found cases of hybrid speciations. \citet{Meng2009} studied four genes in four species of cicadas from New Zealand to support an hypothesized hybrid origin for one species. \citet{Yu2012} and \citet{Bloomquist2010}, using a Maximum Likelihood and a Bayesian approach, investigated 106 genes from yeast species (6 in \citet{Bloomquist2010}, 5 in \citet{Yu2012}) and agreed about their inference of hybridization ancestral to two species. In addition, \citet{Bloomquist2010} studied 9 gene regions in spirochaete Bacteria, and confirmed previous results that one horizontal gene transfer happened in the history of these genes. Thanks to their integrative Bayesian method, they were able to date this event. Finally, \citet{Yu2012} studied more than 9000 genes in three drosophila genomes and also detected hybridization ancestral to one of the three species, this time in disagreement with \citet{Pollard2006}, whose analysis concluded that incomplete lineage sorting was enough to explain the pattern of incongruence in these genomes. Overall, these results show that network-based methods are powerful and can detect past hybridization events. Only \citet{Bloomquist2010}'s method can infer the network topology, but the other methods can be run on a set of topologies to compare their likelihoods.

%In particular, \citet{Szollosi2012}, studying 36 species of Cyanobacteria, found a species tree topology that was perfectly consistent with previous estimates. One difference appeared however in the position of the root, which could be inferred thanks to horizontal gene transfers.

\subsection*{Improving gene tree reconstruction and learning about genome evolution}

	Beyond species tree reconstruction, coalescent models have also been used to investigate genomic evolution in closely related species. For instance, a method based on a Hidden Markov Model was used to estimate divergence times, population sizes and recombination rates in several species of primates. Insights include weaker selection operating on the X chromosome than expected \citep{Hobolth2007}, a negative correlation between chromosome size and chromosome-specific ancestral efficient population size \citep{Mailund2011}, indicative of the power of recombination to increase effective population size, and evidence for selection operating on genes \citep{Hobolth2011}.

	The use of a species tree in addition to a gene alignment yields better gene trees than methods that only consider the gene alignment. \citet{Akerborg2009} studied a dataset of about 180 gene families in 17 yeast genomes with two methods, their own method that uses the sequence alignment and a species tree, and mrBayes, that only uses the alignment \citep{Ronquist2012}. Several of these yeast species descend from a species whose genome has been duplicated. As a consequence all gene trees in the data set must show a duplication event in the branch containing this species. They found that their method detects a branch corresponding to a whole genome duplication in 66\% of the gene families, when mrBayes only detects this branch in 35\% of the cases.
	\citet{Rasmussen2010} obtained a similar result by comparing the inferred orthologs from gene trees obtained using 11 methods to orthologs inferred from synteny information, on a data set of 16 fungi. The seven methods that use the information provided by the species tree were found to outperform the 4 methods that only use the sequence alignments, agreeing with synteny in about $~90\%$ of the cases versus $~60\%$, respectively. Other tests based on a measure of tree balance after a duplication, or based on simulated data all concurred that the information provided by the species tree and interpreted by DL models greatly improves phylogenetic reconstruction. 
	More recently \citet{Mahmudi2013} sample gene trees and reconciliations in an MCMC framework under a DL model and infer duplication and loss rates on a vertebrate tree. Their conclusion is not only that sequence based trees are often wrong, but also that most parsimonious reconciliations of good gene trees are often unprobable.
	Reconciled gene trees have also been used to detect paralogs that originate from whole genome duplications in teleosts \citep{Ouangraoua2011,Howe2013} or at the basis of vertebrates \citep{Makino2010,Affeldt2013} and understand the causes of their maintenance or detect the current traces of these duplications and reconstruct ancestral genomes. They have also been used to study the evolution of metabolism in fungi \citep{Eastwood2011,Floudas2012}. These authors study fungi that digest wood: brown-rot fungi, which digest only cellulose, and white-rot fungi, which digest both cellulose and lignin, the most resistant component in wood. Focusing on a subset of enzymes, and reconciling their gene trees with the species tree, they find that brown-rot fungi are derived white-rot fungi that have lost several important genes. They also infer that white-rot fungi appeared concommitently with the disappearence of coal deposits, and suggest that lignin decay pathways in white-rot fungi may have caused this disappearence.
	
	Although gene tree reconstruction benefits from the input of an accurate species tree, it can probably suffer from the use of an erroneous species tree. Unfortunately, especially as new genomes get sequenced, accurate and well-resolved species trees are not always available. To address such situations and jointly infer the species tree and gene trees, we developed PHYLDOG \citep{Boussau2013}. We tested this program on both simulated and real data and found that both the species tree and the gene trees were reconstructed with good accuracy. Notably, as with the methods that use a given species tree to reconstruct gene trees, we found that our method improved upon methods that only use sequence alignments as input.

\begin{center}
	\begin{figure}[P]
	\centering
		\includegraphics[width=1.\columnwidth]{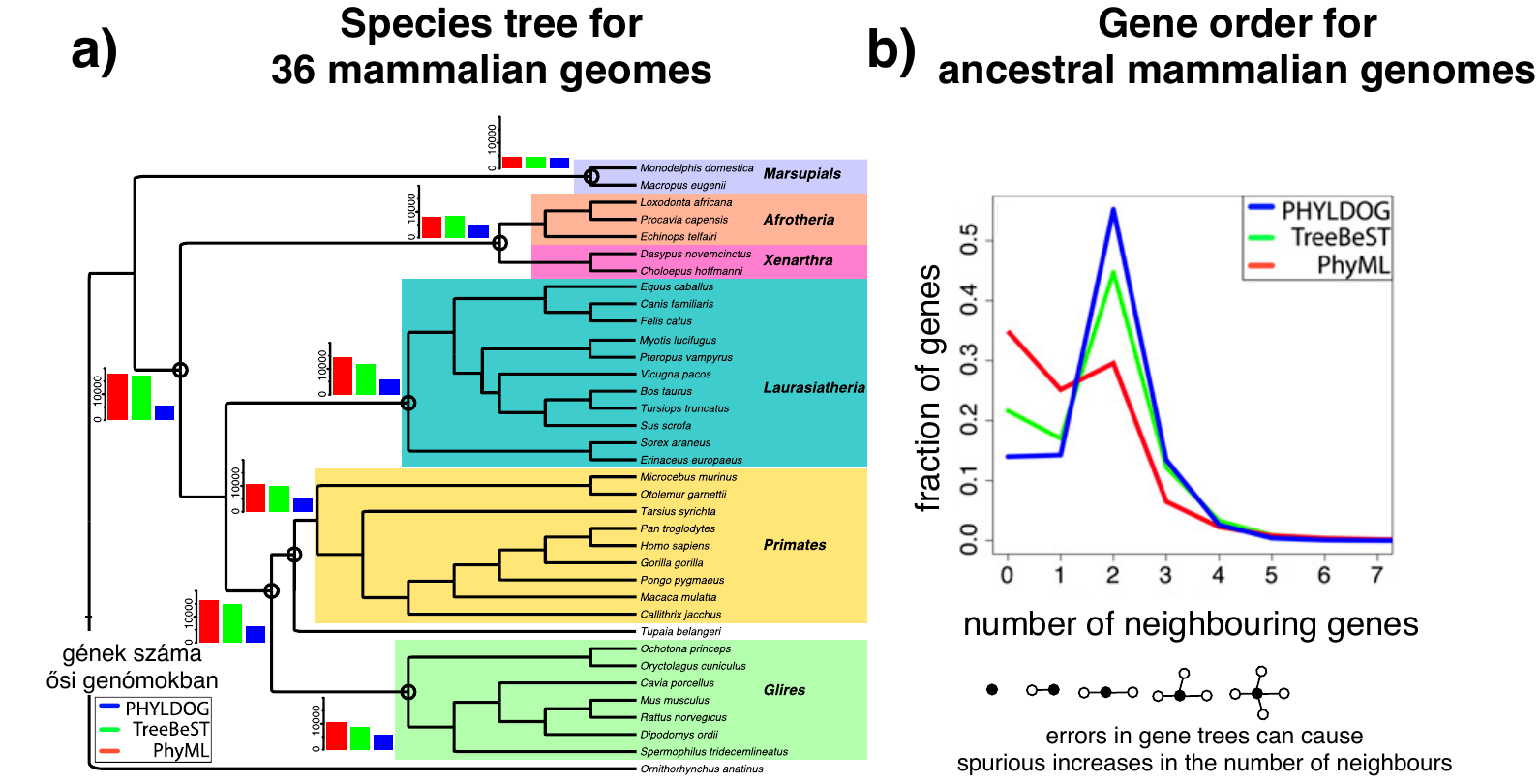}
		\caption{ Left: Species tree inferred by PHYLDOG, with ancestral genome contents reconstructed by different methods on selected nodes. Ancestral genomes reconstructed by PHYLDOG, in blue, have a size similar to that of extant genomes. Right: Reconstruction of ancestral gene syntenies. According to this metric, the method that recovers two neighbors for most genes, as expected from linear chromosomes, is the best. PHYLDOG is in blue.}
		\label{fig:phyldog}
	\end{figure}
\end{center}

%	\subsection*{Inference of a species tree, time orders and genomic evolution with a duplication-loss-transfer model and thousands of gene trees (ODT)}
	Joint reconstruction of the species tree and gene trees is the correct thing to do from a statistical point of view, but is computationally very challenging.  In \citet{Szollosi:2013fk}, we used a novel approach that approaches the accuracy of approaches that infer gene trees while being easier to put in practice. We considered as our data a distribution of gene trees for each gene family instead of just a single gene tree. Using the ALE dynamic programming algorithm (see section "Simulation and inference"), this approach provides a fast but accurate approximation of the actual amount of phylogenetic information contained in the sequence alignment. We found that by not using a single gene tree as our data, the estimates of the number of gene transfers in the Cyanobacterial data set were $~60\%$ lower than when a single gene tree was used. This suggests that the claim that there have been too many transfers in Bacteria and Archaea for reconstructing the tree of life may have been a premature exaggeration. 
	\citet{Szollosi:2013fk} provides a method to reconstruct a gene tree given a species tree and rates of duplication, transfer, and loss, which can be given or all be inferred provided enough information is given to the program. Simulations and measures based on reconstructed ancestral genomes show that these gene trees are more accurate, but the biological relevance of how improved these trees are is perhaps best shown by ancestral sequence reconstruction. Groussin et al. (submitted) reconstructed sequences based on trees inferred through the \citet{Szollosi:2013fk} approach, which uses the species tree and a distribution of gene trees, or through PhyML \citep{Guindon2010a}, an accurate method that does not take the species tree into account. On simulations, this comparison showed that the ancestral sequences were much more accurate when based on the trees obtained with the help of the species tree. More strikingly, the in-vitro resurrection of a protein belonging to the ancestor of Firmicutes, an ancient group of Bacteria, showed that the protein reconstructed based on the method using the species tree was thermodynamically more stable than the protein reconstructed from the alignment-only tree, and had exhibited better enzymatic capabilities. As ancestral sequence resurrection is a popular and powerful approach \citep{Perez-Jimenez2011,Thomson2005,Gaucher2008,Gaucher2003,Bridgham2009,Finnigan2012,Harms2013}, methods using a model of gene family evolution could make an important contribution towards a better understanding of molecular evolution.

	%Reconciled gene trees contribute to better ancestral genomes, better ancestral sequences (Phyldog+ODTL+Groussin, personal communication)

\section*{Next challenges}

Methodological issues are still numerous and leave open wide research avenues, while at the same time the potential
of already available methods can be exploited on an increasingly large scale.

\subsection*{Bypassing the gene tree in the multi-species coalescent }

The multi-species coalescent model describes the evolution of polymorphisms along a species phylogeny. Computing the likelihood of a gene alignment using this model requires summing over a large space of gene trees, given a species tree. This computational difficulty is a major hurdle to using this approach on large data sets, containing large numbers of species, and large numbers of gene families. Very recently, \citet{Bryant2012} and \citet{DeMaio2013} came up with two elegant approaches to computing the likelihood of an alignment under the multi-species coalescent, by bypassing entirely the gene tree level, and instead analytically integrating over the space of possible allele histories. \citet{Bryant2012} consider biallelic data, and provide a model and an algorithm, called SNAPP, that can be used to reconstruct a species tree given an alignment of single nucleotide polymorphisms for instance. They develop a specific algorithm to address the fact that the coalescent process fundamentally functions from the tips of the species tree to its leaves, while the mutation process works forwards. They use this algorithm to reconstruct species trees with 69 individuals in 6 species of Digitalis plants. \citet{DeMaio2013} instead propose a model for sequence data with $A,C,G,T$ data by using a substitution matrix over a larger state space than the usual $4\times4$ substitution matrices: it contains all 6 biallelic states $\{A,C\}$, $\{A,G\}$... with a range of frequencies. They focus on a specific model, where they consider a range of 10 possible frequencies per biallelic frequencies: for the state $\{A,C\}$, we therefore have the states $\{A10\%,C90\%\}$, $\{A20\%,C80\%\}$, ..., $\{A90\%,C10\%\}$. Two approximations are made: first, no more than 2 alleles at a given site can be found at any time in a population, and second their frequencies are well approximated by the limited range included in the model. They construct transitions between states of this matrix from a population size parameter, a selection coefficient, and mutation rates. The resulting instantaneous rate matrix is then exponentiated to provide a matrix of substitution probabilities. Overall, the matrix obtained with a range of 10 possible frequencies per biallelic state contains 58 states, \textit{i.e.} about the same number of states as a codon substitution model. \citet{DeMaio2013} use this model, with some further refinements to account for context-dependent mutations and strand-specificity on a large alignment of four species of primates and find evidence for a smaller ancestral population size in orangutans, and selection on splicing enhancers in exons.

Such analytical approaches seem very promising for combining coalescent models with duplication, loss and transfer models, as they bypass the problem of sampling allele histories. How they improve upon multispecies coalescent gene tree-species tree models is still an open question.

\subsection*{More Integrative models}

The integrative program of \citet{Goodman1979} is being progressively implemented. The probabilistic framework makes it natural to integrate mutations with duplications and losses through the coalescent \citep{Rasmussen2012}, duplications, losses and transfers with substitutions \citep{Boussau2013,Szollosi2012,Szollosi2013,Szollosi:2013fk}. Rearrangements can be handled in parsimony if ILS is ignored \citep{Berard2012,Patterson2013}.

%Integration of several levels of genome evolution has now reached the point where it is possible to handle
%together ILS, DL and sequence-based evolution \citep{Rasmussen2012}, adding the possibility
%of transfers \citep{Stolzer2012} and additionally rearrangements if ILS is ignored \citep{Berard2012,Patterson2013}
%in parsimony frameworks. Gene trees and species trees are
%co-estimated in coalescent \citep{Heled2010,Liu2008}, or DL \citep{Boussau2013} probabilistic frameworks.
%, or DTL \citep{Szollosi2012,Szollosi2013,Szollosi:2013fk}.

A model and method to handle a union of all of these processes is currently missing.
However there are very good reasons for the integration of different levels of data analysis to continue. For instance, below the gene tree / species tree problem, is the inference of gene alignments. Only recently has the problem of joint inference of alignments and gene trees been considered seriously, with attempts to model the process of insertion/deletion in the evolution of sequences. Such approaches show dramatic improvements over phylogenetically unaware alignment methods \citep{Redelings2005,Satija2009,Warnow2013}. 
However, they obviously need all the information necessary to have the best possible gene tree, \textit{e.g.} a link to the species tree. Hence, it is probable that the integration of gene tree-species tree models and alignment methods should benefit to the inference of alignments, gene trees and perhaps species trees.

Although a global model seems difficult to imagine presently, the entire pipeline of sequence data analysis, from sequencing error corrections to gene annotation and genome assembly is likely to benefit from probabilistic evolutionary models.  The recognition of homologous sequences, the prediction of gene functions based on information from other organisms, and the proximity of genes on chromosomes all depend ultimately on the structure of the species tree and the possible events of substitution, duplication, loss and lateral transfer that may have occurred in the history of genomes. There is currently no proposition of an integration of these processes on all levels of the pipeline described in Fig. 5, but phylogenetically aware methods have proved very promising at many different steps of the process \citep{Boussau2010} including on genome assembly \citep{Husemann2010,Rajaraman2013}.

\subsection*{Algorithmics and computing time}

The score of a gene tree, especially if it is the combination of scores from several models,
can be fairly costly to compute. Therefore the exploration of trees is always time consuming. Already the inference of a gene tree that maximizes the probability of the alignment given the gene tree is provably hard. The joint inference, estimation of parameters and exploration
of dated or ordered species trees cumulate intractable problems.
In practice optimizing a gene tree can necessitate up to a few hours for very large families. As
there can be thousands of gene families in a typical dataset, the computations even for a fixed species tree can take
a long time.
However models of gene family evolution as well as sequence-based models all make the assumption that genes evolve
independently from each other. This assumption can be questioned (see below)
and is also broken by evolutionary parameters shared among gene families. But it allows a trivial parallelization by the data. All genes trees can be computed independently, given a common species tree.
Hence, a species tree exploration is mainly constrained by the largest multigene families. A simple way to increase computational efficiency is to ignore these large families in a first step of species tree exploration. Large multigene families can be considered later, when a good species tree is found based on smaller gene families, or, in a sampling context, using importance sampling. However, such tricks can only help as long as the number of genomes under study is relatively small. For studying larger datasets, we will need to devise more efficient algorithms.

\subsection*{Reconstructing and dating the tree of life}

The claim that Darwin, who failed to foresee the role of lateral gene transfer in microbial evolution, was wrong in the few texts in which he put forward the metaphor of a tree of life is arguably the result of a confusion between gene trees and species trees \citep{Doolittle1999}. The models we have described here actually show that the plurality of gene histories can not only be overcome but more importantly provide additional information on the processes and patterns of evolution. The phylogenies for a diversity of clades have been reconstructed with coalescent, DL or DTL models. In each case, the degree of conflict among gene trees can be interpreted in biological terms, such as divergence time and ancestral population size with the coalescent, or relative timing of speciation with LGT. There is a great hope that the development and use of these models will help resolve many issues that were left pending by traditional methods. 

\subsection*{Beyond the gene as an evolutionary unit}

\begin{center}
	\begin{figure}[P]
	\centering
		\includegraphics[width=1.\columnwidth]{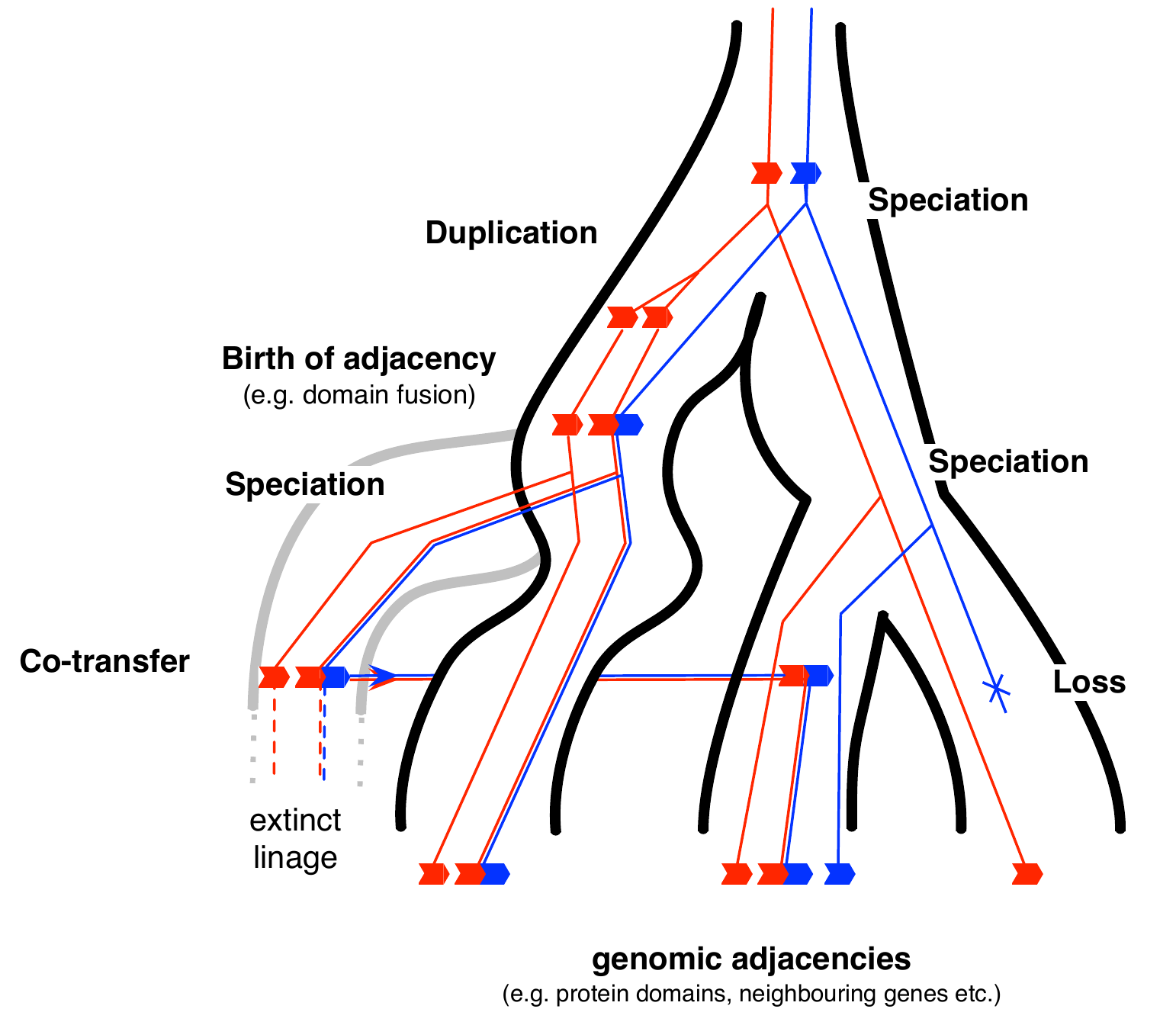}
		\caption{ Evolutionary units below or above genes. Red and blue units can be domains inside genes
				or genes that are neighbours along a chromosome or genes involved in a protein complex. Adjacencies are binary relations between genes, and evolve along a species phylogeny. Adjacencies can be gained or lost regardless of the birth and death of the units. When two units together undergo speciation, duplication or transfer, adjacencies undergo the same events.}
		\label{fig:murray}
	\end{figure}
\end{center}

Although we have adopted a liberal sense for "gene" (any genomic sequence can be called a gene), in many of the studies we reported, a gene is a sequence coding for a protein or a functional RNA, and is considered as an evolutionary unit. However, within such genes, different parts may have different histories \citep{Didelot2010,Wu2012}. Alternatively, some genes may be associated throughout evolutionary times because their functions are interdependent or simply because they are close to each others in the genome. As such, they may be duplicated or transferred together \citep{Bansal2013,Patterson2013}. Hence, the definition of evolutionary units is difficult, and fluctuates in time (see Figure \ref{fig:murray}).
As we have shown, almost all existing models describe the reconciliation of one gene tree with one species tree, supposing its evolution is coherent and independent from other genes. Some genomic studies however use branchwise but genome wide evolutionary parameters like the rate of duplications and losses \citep{Boussau2013}. This can be seen as a trick to model large scale events like genome duplications without doing away with the independence of genes, which is computationally advantageous. But it fails to model more local rearrangements such as duplications of parts of a chromosome. These events could be informative for phylogeny, but models of genome rearrangements are often combinatorially so complex \citep{Fertin2009} that they do not scale up well with the size and number of genomes \citep{York2002,Darling2008,Miklos2010}. Until now, their complexity has precluded a coupling with other models such as gene tree - species tree reconciliation. However, assuming neighborhoods between genes are independent (but genes are not), it is possible to integrate rearrangements into DL \citep{Berard2012} or DTL \citep{Patterson2013} models. Such approaches describe the evolution of neighborhoods (or any other relationship between genes, including functional ones) along pairs of reconciled gene trees, allowing one to reconstruct adjacencies in ancestral genomes and evolutionary events of duplication, loss and transfer that have affected genomic fragments comprising several genes. Because such multiple events are frequent, it is likely that the parameters of duplication, transfer and loss that are estimated in DL and DTL models are biased and it seems necessary to integrate models of neighborhood evolution with phylogenetic reconstruction into the reconstruction of genome histories.

There are also models for detecting breakpoints inside gene sequences using HMMs for instance \citep{McGuire2000,Suchard2002,Martins2008,Boussau2009a}, or detecting breakpoints of phylogenetic discordance at a whole genome scale \citep{Ane2011}, but so far these models have not been included in models of gene family evolution.

\subsection*{Keeping up with the pace of data acquisition}

Currently genome sequencing is no longer a limiting step for comparative genomics. Instead, assembling gene families, gene alignments, gene trees and a species tree are becoming increasingly  problematic. In this context, methods using models of gene family evolution may have their advantage because they effectively reduce the space of possible solutions to explore: given a species tree, the space of possible gene trees is limited compared to species tree unaware methods, and, consequently, the space of possible alignments. Devising smart algorithms, which make use of these reductions of complexity may provide fast yet accurate inferences for large scale comparative genomics projects. 

Another area where progress is needed is in the reuse of prior information. Currently, every time a new comparative genomics project is undertaken, or every time a database of homologous sequences is updated, many inference tasks need to be redone from scratch. The computations of gene families, alignments, trees and species tree are usually done as if there was no prior information obtained from previous analyses. This is obviously a huge waste of useful information, as these computations are often very demanding. Future approaches to comparative genomics will need to be not only integrative, but also incremental. There is a large need for new developments, and the Systematic Biology community is well equipped to undertake them.

\section*{Conclusion}

In the past fifteen years, the relationship between gene trees and the species tree has been greatly clarified. This conceptual advance has been accompanied by methodological developments in models of gene family evolution and in the algorithms needed for statistical inference. These rely heavily on birth-death processes and dynamic programming. In the next few years, these developments will go two seemingly incompatible ways: they will have to increase in complexity to more accurately model genome evolution, but they will also need to scale up as data sets keep increasing. This tension presents exciting challenges.

%\begin{acknowledgments}
%We thank all members of the Bioinformatics and Evolutionary Genomics Group for discussions of the results and comments on the manuscript.
This project was supported by the French Agence Nationale de la Recherche (ANR) through Grant ANR-10-BINF-01-01 ``Ancestrome''.
 GJSz was supported by the Marie Curie CIG 618438 ``Genestory'' and the Albert Szent-Gy\"orgyi Call-Home Researcher Scholarship A1-SZGYA-FOK-13-0005.
%\end{acknowledgments}

%\bibliographystyle{sysbio}
% \bibliographystyle{sysbio}
% \bibpunct[; ]{(}{)}{;}{a}{}{;}

 \bibliographystyle{plainnat}
%\bibliography{biblioClean,biblio2Clean,biblio_bibdesk}
\bibliography{biblio}

\end{document}